\documentclass[times,twocolumn]{aastex631}

\usepackage{graphicx}

\usepackage{enumitem}
\usepackage{amsmath}
\usepackage{multirow}
\usepackage{booktabs}
\usepackage{array}
\usepackage{bm}

\usepackage{array}
\newcolumntype{x}[1]{>{\centering\arraybackslash\hspace{0pt}}p{#1}}

\usepackage{CJK}

\usepackage{etoolbox}
\makeatletter
\patchcmd{\NAT@citex}
  {\@citea\NAT@hyper@{%
     \NAT@nmfmt{\NAT@nm}%
     \hyper@natlinkbreak{\NAT@aysep\NAT@spacechar}{\@citeb\@extra@b@citeb}%
     \NAT@date}}
  {\@citea\NAT@nmfmt{\NAT@nm}%
   \NAT@aysep\NAT@spacechar\NAT@hyper@{\NAT@date}}{}{}
\patchcmd{\NAT@citex}
  {\@citea\NAT@hyper@{%
     \NAT@nmfmt{\NAT@nm}%
     \hyper@natlinkbreak{\NAT@spacechar\NAT@@open\if*#1*\else#1\NAT@spacechar\fi}%
       {\@citeb\@extra@b@citeb}%
     \NAT@date}}
  {\@citea\NAT@nmfmt{\NAT@nm}%
   \NAT@spacechar\NAT@@open\if*#1*\else#1\NAT@spacechar\fi\NAT@hyper@{\NAT@date}}
  {}{}
\makeatother

\received{January 25, 2021}
\revised{March 11, 2021}
\accepted{March 13, 2021}

\shorttitle{Emerging Dimming}
\shortauthors{Payne and Sun}

\begin{document}

\begin{CJK*}{UTF8}{gbsn}

\title{Emerging Dimming as Coronal Heating Episodes}

\author[0000-0003-3490-3243]{Anna V. Payne}
\altaffiliation{NASA Graduate Fellow}
\affil{Institute for Astronomy, University of Hawai`i at M\={a}noa, Honolulu, HI 96822, USA}

\author[0000-0003-4043-616X]{Xudong Sun (孙旭东)}
\affil{Institute for Astronomy, University of Hawai`i at M\={a}noa, Pukalani, HI 96768, USA; \href{mailto:xudongs@hawaii.edu}{xudongs@hawaii.edu}}



\begin{abstract}
Emerging dimming occurs in isolated solar active regions (ARs) during the early stages of magnetic flux emergence. Observed by the Atmospheric Imaging Assembly, it features a rapid decrease in extreme-ultraviolet (EUV) emission in the 171~{\AA} channel images, and a simultaneous increase in the 211~{\AA} images. Here, we analyze the coronal thermodynamic and magnetic properties to probe its physical origin. We calculate the time-dependent differential emission measures for a sample of 18 events between 2010 and 2012. The emission measure (EM) decrease in the temperature range $5.7 \le \log_{10}T \le 5.9$ is well correlated with the EM increase in $6.2 \le \log_{10}T \le 6.4$ over eight orders of magnitude. This suggests that the coronal plasma is being heated from the quiet-Sun, sub-MK temperature to 1--2 MK, more typical for ARs. Potential field extrapolation indicates significant change in the local magnetic connectivity: the dimming region is now linked to the newly emerged flux via longer loops. We conclude that emerging dimming is likely caused by coronal heating episodes, powered by reconnection between the emerging and the ambient magnetic fields.
\end{abstract}
 


\section{Introduction}
\label{sec:intro}

Our nearest star, the Sun, serves as a unique laboratory for stellar processes. In particular, strong magnetic fields in active regions (ARs) interact with the coronal plasma, changing its thermal structure as they emerge.  Studying the formation, evolution, and decay of ARs allows for a deeper understanding of solar variability, and how it impacts us on Earth in the form of space weather. The relation between the magnetic field and the plasma thermodynamic properties may hold the key to solving the coronal heating problem.

\begin{figure}[t!]
\centerline{\includegraphics[width=\columnwidth]{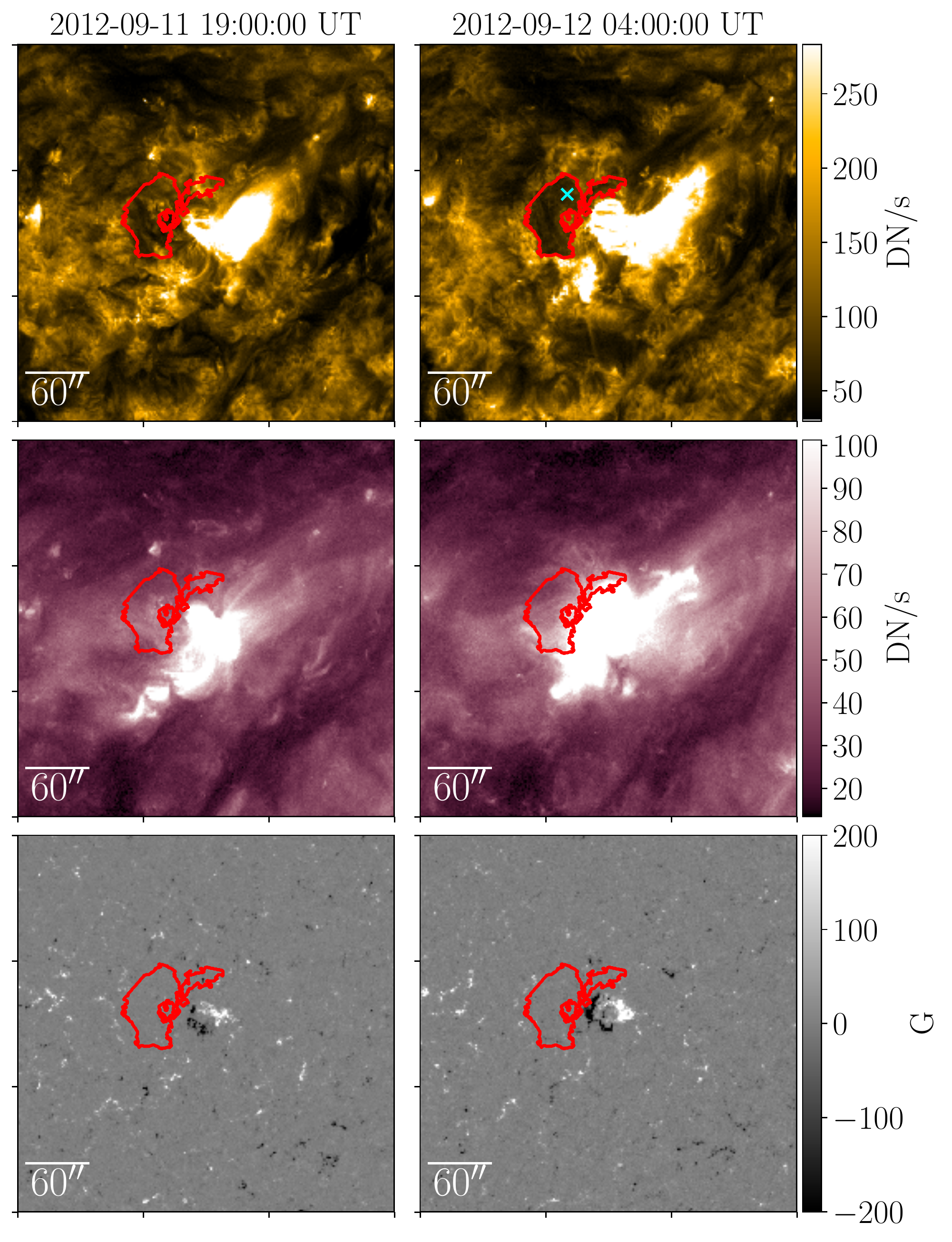}}
\caption{An example of emerging dimming around AR 11570, which is used as a representative example throughout the paper. Left and right columns show \textit{SDO} observations taken at the start and the maximum of dimming, about 9~hr apart. Top and middle rows show AIA 171 and 211~{\AA} channels; bottom row shows HMI LOS magnetograms. The red contours outline the emerging dimming region. The cyan cross denotes the location of the pixel used in Figure~\ref{fig:dem} as an example.}
\label{fig:sdoaia}
\end{figure}

One interesting feature recently discovered is the ``emerging dimming'' \citep{zhang2012} in isolated ARs. This type of ARs emerge with no other pre-existing AR in the vicinity, so the local magnetic flux is likely balanced with little open flux. As initially reported, 24 isolated ARs between 2010 June to 2011 May exhibited decreases of emission in an extreme-ultraviolet (EUV) channel dominated by the lower-temperature \ion{Fe}{9} 171~{\AA} line (0.6~MK) during the early emerging stages. For the higher-temperature lines, in particular \ion{Fe}{14} 211~{\AA} (2~MK), the emission increased continuously (Figure~\ref{fig:sdoaia}). No dimming was observed in other channels. The emerging dimming regions are situated \textit{next to} or \textit{around} the emerging flux, with a fan or halo shape, respectively. \citet{zhang2012} speculated that coronal magnetic reconnection between the emerging and background fields heats up the coronal plasma, thus causing the cooler channel to dim and the warmer channels to brighten.

Here we analyze the coronal thermodynamic and magnetic properties of emerging dimming. We specifically test the hypothesis that the observed dimming is caused by plasma heating, rather than other mechanisms such as mass loss. Using multi-wavelength EUV data, we calculate the time-dependent differential emission measure (DEM) and the emission measure (EM) in selective temperature ranges. Further, we study the change of the magnetic field connectivity for a representative case using a potential field extrapolation model. These calculations yield insights on how the density of the optically-thin plasma at different temperatures changes temporally and in response to the surface magnetic field, thus allowing us to probe the underlying physical processes.


\section{Data \& Methods}
\label{sec:methods}


\subsection{Observation}
\label{subsec:observation}

With the advent of the \textit{Solar Dynamics Observatory} \citep[\textit{SDO};][]{pesnell2012}, launched in 2010, the solar corona is continuously monitored over 10 UV/EUV wavelengths with a cadence of 12~s and spatial sampling of 0$\farcs$6 with the Atmospheric Imaging Assembly \citep[AIA;][]{lemen2012}. In addition, the Helioseismic and Magnetic Imager \citep[HMI;][]{schou2012}, also on board \textit{SDO}, measures both line-of-sight (LOS) and vector magnetic fields in the photosphere. The LOS and vector magnetograms are taken at a cadence of 45~s and 12 minute respectively, with a spatial sampling of 0$\farcs$5.

With high cadence and a continuous, voluminous dataset, \textit{SDO} is a rich resource for studying how emerging dimming forms, and how they change through time. By observing the Sun in different, optically-thin EUV channels with AIA, information about the coronal electron density and temperature is recorded through time. Contemporaneous EUV and magnetic field observations allows for direct association between the coronal dynamics and their photospheric sources. 

We used the list of emerging ARs reported in \citet{schunker2016} to search for cases of emerging dimming. The two main properties provided by the list are (1) their visibility in the continuum, and (2) the degree of spatial isolation, i.e., whether the emergence occurs in relatively quiet areas of the Sun that do not contain, and are not in the vicinity of, pre-existing ARs. The second criterion allows us to select only isolated ARs where emerging dimming was first identified \citep{zhang2012}. It also avoids the possible influence of AR-AR interaction so the interpretation of the results is more straightforward.


\subsection{Data Acquisition \& Processing}
\label{subsec:data}

We search for emerging dimming events over a two-year period from 2010 to 2012 based on visual inspection. Using the starting times of flux emergence given in \citet{schunker2016}, we check the AIA 171~{\AA} images using \textit{JHelioviewer} (\citealt{muller2017}), which provides an easy way to query and visualize \textit{SDO} data. If we observe significant dimming next to or around the emerging flux, we manually define (1) a time range (typically a few days) to encapsulate the full history of the AR, including times before and during full emergence, and (2) a fixed field of view (FOV) in the native Helioprojective coordinate. We subsequently extract \textit{SDO} data cubes corresponding to this time range and FOV for each event at 30~minute cadence. The data cubes include six AIA EUV channels (94, 131, 171, 193, 211, and 335~{\AA}) and HMI LOS magnetograms. Our final sample consists of 18 events (Table~\ref{table:dimmingtable}): nine were included in \citet{zhang2012}; nine are new from 2012. We discuss the efficacy of this sample in Section~\ref{sec:discussion}.

We perform all data reductions using the SolarSoftWare (SSW) IDL package. SSW query of the \textit{SDO} database returns level-1 AIA images, in which the raw images have been de-spiked, had bad pixels removed, and flat-fielded. We use the module \texttt{aia\_prep.pro} to process the images into level 1.5, adjusting images to a common plate scale and ensuring all images are centered at the same pixel. The images are also de-rotated and co-aligned so as to enable a direct comparison between different times.

Our analysis focuses on an ``emerging dimming region'' defined as a contiguous, enclosed area whose 171~{\AA} brightness declined post-emergence of the central AR. The contours for each AR were selected based on the 171~{\AA} image pixel values during times when the AR's surrounding area visibly darkened. This subregion is fixed for each AR through its entire time series. This is illustrated by the red contours in Figure~\ref{fig:sdoaia} for a sample event in AR 11570. The dimming regions were extracted based on a contour applied to the 171~{\AA} images after the time of emergence given in \citet{schunker2016}. The contour threshold was set to be below half of the average surrounding medium around the AR in the 171~{\AA} data to form a contiguous entity surrounding the AR. The same contour was then applied to the 211~{\AA} images and magnetograms. We consider the start of the dimming time, $t_0$, around the time of emergence given in \citet{schunker2016}. The time of maximum dimming, $t_{*}$, was then determined as the minimum of the spatially integrated EM$_{\mathrm{low}}$ curve, described in the following section.  


\begin{table*}[t!]
\begin{center}
\caption{Thermodynamic properties of all 18 emerging dimming events analyzed in this work.\label{table:dimmingtable}}
\begin{tabular}{cccccx{1.9cm}x{1.9cm}x{1.3cm}x{1.3cm}c} 
\toprule 
\multirow{2}{*}{AR} & \multirow{2}{*}{$t_{*}$} & \multirow{2}{*}{$t_{*}-t_0$~(hr)} & \multicolumn{2}{c}{$I(t_{*})/I_\mathrm{QS}$} & \multicolumn{2}{c}{$\sum\mathrm{EM}(t_{*}) - \sum\mathrm{EM}(t_0)$~(cm$^{-3}$)} & \multicolumn{2}{c}{$\sum\mathrm{EM}(t_{*})/\sum\mathrm{EM}(t_0)$} & Morphology\\
 &  &  & 171~{\AA} & 211~{\AA} & low & high & low~ & high \\
\midrule
11122 & 2010-11-06T05:00 & 17.0 & 0.50 & 2.19 & $-7.14\times10^{55}$ & $9.58\times10^{56}$ & 0.06 & 1.66 & Fan \\
11179 & 2011-03-21T12:30 & 16.5 & 0.50 & 2.61 & $-2.56\times10^{51}$ & $5.15\times10^{52}$ & 0.02 & 7.55 & Fan \\
11194 & 2011-04-13T04:30 & 6.5 & 0.75 & 3.23 & $-2.45\times10^{49}$ & $2.20\times10^{51}$ & 0.06 & 2.21 & Fan \\
11198 & 2011-04-22T06:30 & 7.0 & 0.68 & 3.90 & $-1.54\times10^{51}$ & $3.33\times10^{52}$ & 0.09 & 1.34 & Halo \\
11211 & 2011-05-08T03:30 & 5.5 & 0.74 & 0.68 & $-7.73\times10^{51}$ & $9.11\times10^{52}$ & 0.24 & 1.20 & Fan \\
11214 & 2011-05-13T23:12 & 10.5 & 0.55 & 1.15 & $-2.50\times10^{51}$ & $9.75\times10^{51}$ & 0.28 & 1.01 & Fan \\
11215 & 2011-05-12T04:30 & 13.5 & 0.53 & 1.33 & $-1.07\times10^{55}$ & $5.09\times10^{55}$ & 0.19 & 1.12 & Fan\\
11220 & 2011-05-22T05:00 & 11.0 & 1.16 & 2.32 & $-4.82\times10^{53}$ & $1.57\times10^{55}$ & 0.11 & 1.87 & Fan \\
11221 & 2011-05-22T03:30 & 3.5 & 0.80 & 1.46 & $-1.06\times10^{50}$ & $1.66\times10^{52}$ & 0.71 & 1.08 & Fan \\
\midrule
11400 & 2012-01-13T22:48 & 14.0 & 0.47 & 1.25 & $-9.94\times10^{54}$ & $1.68\times10^{56}$ & 0.16 & 1.22 & Fan \\
11414 & 2012-02-04T11:24 & 10.0 & 0.58 & 1.33 & $-3.29\times10^{54}$ & $1.30\times10^{56}$ & 0.12 & 1.47 & Fan \\
11431 & 2012-03-04T15:48 & 15.0 & 0.73 & 2.62 & $-8.00\times10^{53}$ & $1.32\times10^{55}$ & 0.02 & 4.43 & Fan \\
11437 & 2012-03-16T14:48 & 7.0 & 0.63 & 1.17 & $-5.69\times10^{55}$ & $1.17\times10^{57}$ & 0.06 & 1.65 & Halo \\
11446 & 2012-03-22T22:24 & 10.0 & 0.53 & 1.74 & $-2.80\times10^{54}$ & $3.04\times10^{56}$ & 0.25 & 1.25 & Fan \\
11570 & 2012-09-12T04:00 & 9.0 & 0.73 & 1.94 & $-4.27\times10^{54}$ & $8.55\times10^{56}$ & 0.27 & 1.38 & Fan \\
11603 & 2012-10-31T00:48 & 12.0 & 0.64 & 2.96 & $-2.13\times10^{54}$ & $1.11\times10^{56}$ & 0.16 & 1.35 & Fan \\
11607 & 2012-11-04T22:18 & 5.5 & 0.59 & 1.92 & $-3.03\times10^{56}$ & $2.33\times10^{57}$ & 0.11 & 1.23 & Fan \\
11624 & 2012-11-28T04:00 & 16.0 & 0.74 & 1.42 & $-4.88\times10^{55}$ & $1.38\times10^{57}$ & 0.25 & 1.36 & Fan \\
\bottomrule 
\end{tabular}\\ [2mm]
\parbox[c]{\textwidth}{
{\textbf{Notes.}} 
The first nine events were included in the original sample in \citet{zhang2012}; the last nine are new cases. $t_{*}$ and  $t_0$ indicate the time of maximum dimming and the beginning of flux emergence, respectively. $I(t_{*})/I_\mathrm{QS}$ is the average brightness within the dimming contour at $t_{*}$ divided by the average brightness of the nearby quiet Sun (QS). $\sum\mathrm{EM}(t_{*}) - \sum\mathrm{EM}(t_0)$ is the change of the EM integrated over the emerging dimming region; $\sum\mathrm{EM}(t_{*})/\sum\mathrm{EM}(t_0)$ shows the relative change. The ``low'' and ``high'' columns refer to EM calculated for the temperature range $5.7 \le \log_{10}T \le 5.9$ and $6.1 \le \log_{10}T \le 6.3$, respectively. Random errors of all variables are negligible. Morphology refers to whether the emerging dimming region sits next to the emerging magnetic flux on one side (fan), or fully surround the flux (halo) as classified by \citet{zhang2012}.
} \\
\end{center}
\end{table*}



\begin{figure}[t!]
\centerline{\includegraphics[width=0.85\columnwidth]{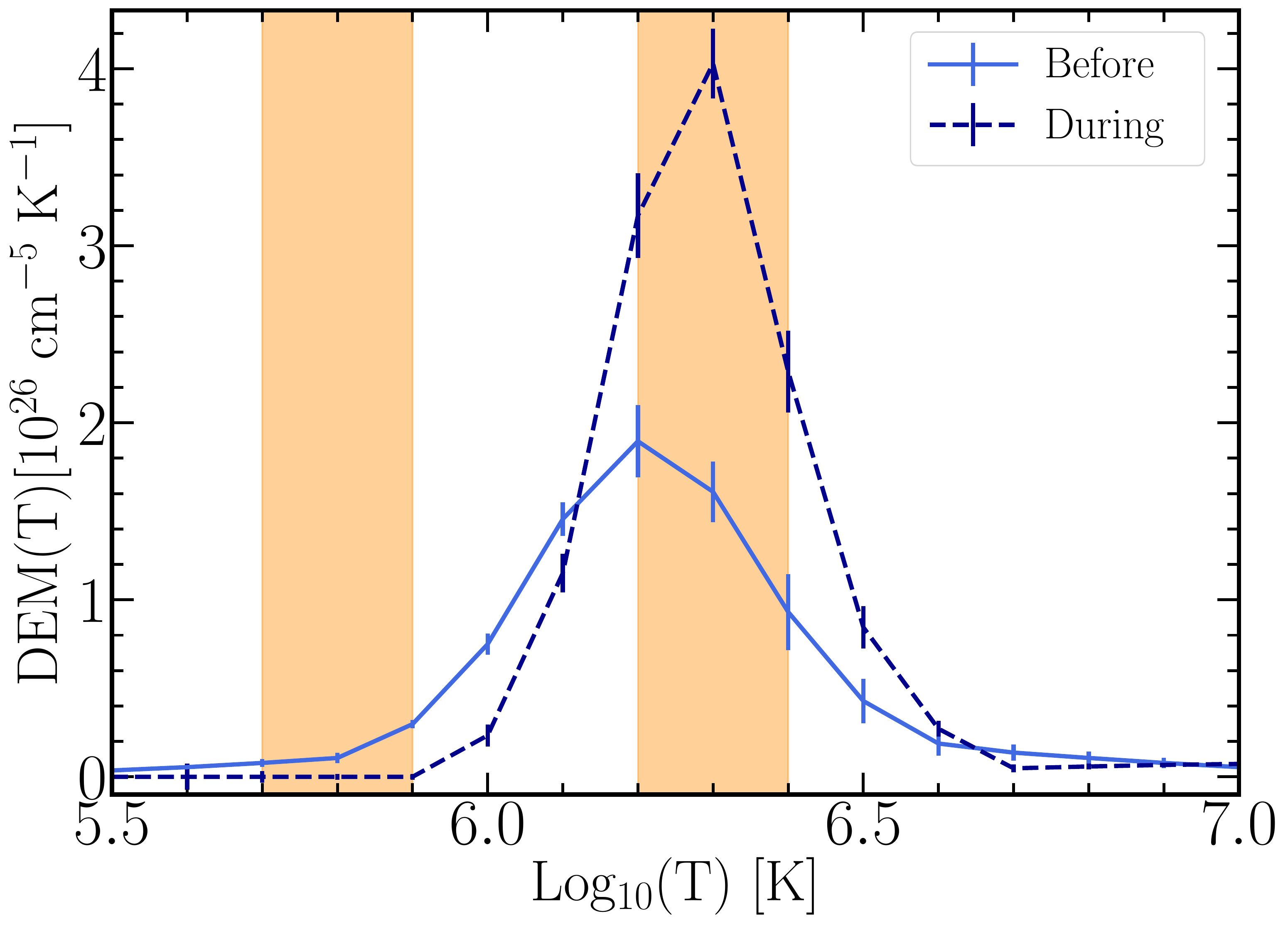}}
\caption{Example DEM solution before (light blue, solid line) and during emerging dimming (dark blue, dashed line) for AR11570. This DEM solution is obtained from one pixel within the map shown in Figure~\ref{fig:sdoaia}. Error bars are caused by random noise, determined by the method of Monte-Carlo resampling. The two transparent orange regions denote the temperature bins used for the EM calculations, where the low temperature bin ranges from $\log_{10}T = 5.7 - 5.9$ and the high temperature bin ranges from $\log_{10}T = 6.2 - 6.4$.}
\label{fig:dem}
\end{figure}


\subsection{Differential Emission Measure (DEM) Analysis}
\label{subsec:dem}

To understand the physical origin of emerging dimming, we analyze the coronal DEM:
\begin{equation}
\mathrm{DEM}(T)\,dT = \int^{\infty}_0 n_e^2(T,z)\,dz.
\label{eq:dem}
\end{equation}
Here, DEM is a function of the temperature $T$; $n_e(T,z)$ is the electron number density as function of $T$ and the spatial coordinate $z$. By convention, $z$ is $0$ at the coronal base and increases toward the observer along the LOS. The integral of DEM over a finite temperature range $T_0 \le T \le T_1$ is called the emission measure (EM):
\begin{equation}
\mathrm{EM}(T) = \int^{T_1}_{T_0} \mathrm{DEM}(T)\,dT.
\label{eq:em}
\end{equation}
The DEM is related to the narrow-band, optically-thin EUV observations by an integral in the temperature space:
\begin{equation}
y_i = \int^{\infty}_0 K_i(T)\,\mathrm{DEM}(T)\,dT,
\label{eq:yi}
\end{equation}
where $y_i$ is the exposure-normalized pixel value (i.e., count rate) in the \textit{i}-th EUV channel, and $K_i(T)$ is the known instrument- and channel-dependent temperature response function. Given a set of AIA observations $y_i$, our goal is to solve for DEM from Equation~\ref{eq:yi} so as to learn about the thermodynamic parameters $n_e$ and $T$. This process is called DEM inversion.

We employ the DEM algorithm described in \citet{cheung2015}, which computes DEM solutions using a linear programming (also called linear optimization) approach based on the concept of sparsity. This differs from other commonly used procedures, which are mostly based on $\chi^2$-minimization. The sparsity constraint reduces the risk of overfitting for underdetermined systems.

After creating data cubes for each emerging dimming event in the six AIA channels, we use the SSW module \texttt{aia\_sparse\_em\_solve.pro} to calculate the DEM solutions over time. In the end, for each event we have a cube of a DEM solution for each AIA pixel, aligned over time to cover the whole duration. The calculations are typically performed in the logarithm temperature space with a $\Delta\log_{10}T=0.1$ bin size.

Our analysis is mostly focused on the total EM integrated over the emerging dimming region, denoted as $\sum\mathrm{EM}$. We integrate for following two temperature ranges: $\sum\mathrm{EM_{low}}$ for $5.7 \le \log_{10}T \le 5.9$, and $\sum\mathrm{EM_{high}}$ for $6.1 \le \log_{10}T \le 6.3$. These ranges are chosen to encapsulate the characteristic temperatures, around the peak of the response function of the 171 and 211~{\AA} channels. 


\begin{figure}[t!]
\centerline{\includegraphics[width=0.85\columnwidth]{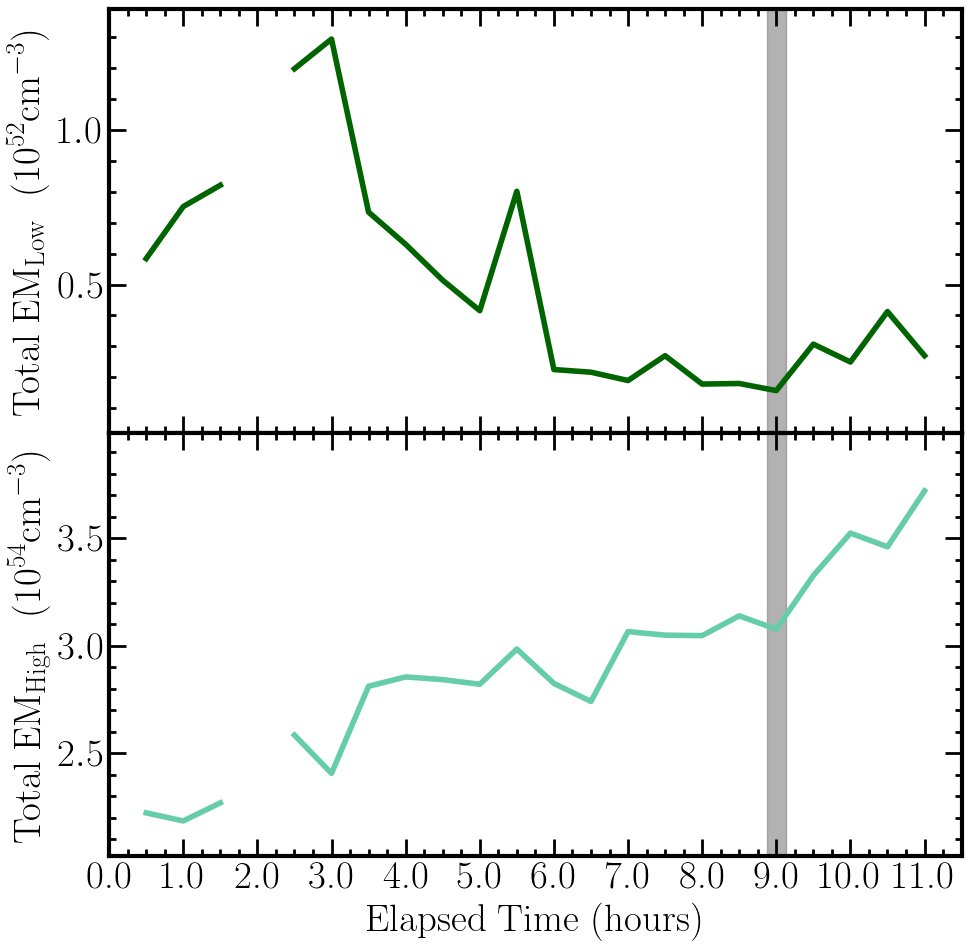}}
\caption{Spatially integrated EM for AR 11570 over the emerging dimming region over time, elapsed from 2012-09-11T19:00 UT which corresponds to the emergence time of the AR. Top: $\mathrm{EM_{low}}$; bottom: $\mathrm{EM_{high}}$.  The transparent vertical line denotes the time of maximum dimming. The gap in the curves is due to missing AIA data. Error bars are too small to be visible.}
\label{fig:dem_compare}
\end{figure}


The algorithm from \citet{cheung2015} does not include a method to estimate the random error associated with the noise in EUV images. In order to estimate the error bars, we devise a method of resampling in a Monte-Carlo-like manner. We utilize the SSWIDL module \texttt{aia\_bp\_estimate\_error.pro} for an estimate of the uncertainty $\sigma(y_i)$ for each nominal AIA pixel value $y_i$ in the $i$-th channel.  Assuming the noise is Gaussian like, we create for each pixel a sample following the normal distribution $\mathcal{N}(y_i, \sigma^2(y_i))$, and repeat for all pixels independently. For a sample size of $N$, this effectively creates $N$ realizations of the same AIA image; this is done for each channel at each time step. After calculating $N$ unique DEM solutions, we estimate the error as the 68\% range centered at the medium of the returned DEM and EM values. We find a sample size $N=100$ returns sufficiently stable quotes of the errors. Due to a large number of contributing pixels, the random error in $\sum\mathrm{EM}$ is negligible. The overall uncertainty is expected to be dominated by the poorly understood systematics, which is on the order of $20\%$ \citep{judge2010}.


\subsection{Potential Field Extrapolation}
\label{subsec:pf}

\citet{zhang2012} hypothesized that coronal magnetic reconnection occurs between the emerging and background fields during emerging dimming events. If so, magnetic connectivity should change, that is, field lines originate from inside emerging dimming regions should have different end points after the AR emergence. To quantitatively assess the changes, we use a local potential field extrapolation algorithm to model the coronal magnetic field. The algorithm is based on a Green's function method \citep{sakurai1989} and is implemented by \citet{wiegelmann2004}.

We use the photospheric magnetic field in the radial direction ($B_r$) derived from HMI vector magnetograms \citep{hoeksema2014} as the lower boundary condition. We ignore the local curvature and perform the extrapolation in a Cartesian coordinate with a 364~km resolution. We subsequently trace field lines from the emerging dimming regions identified in co-aligned AIA images. Field lines with starting points close to the side boundary are not included as to minimize the boundary effect.


\begin{figure*}
    \centering
    \includegraphics[width=0.95\textwidth]{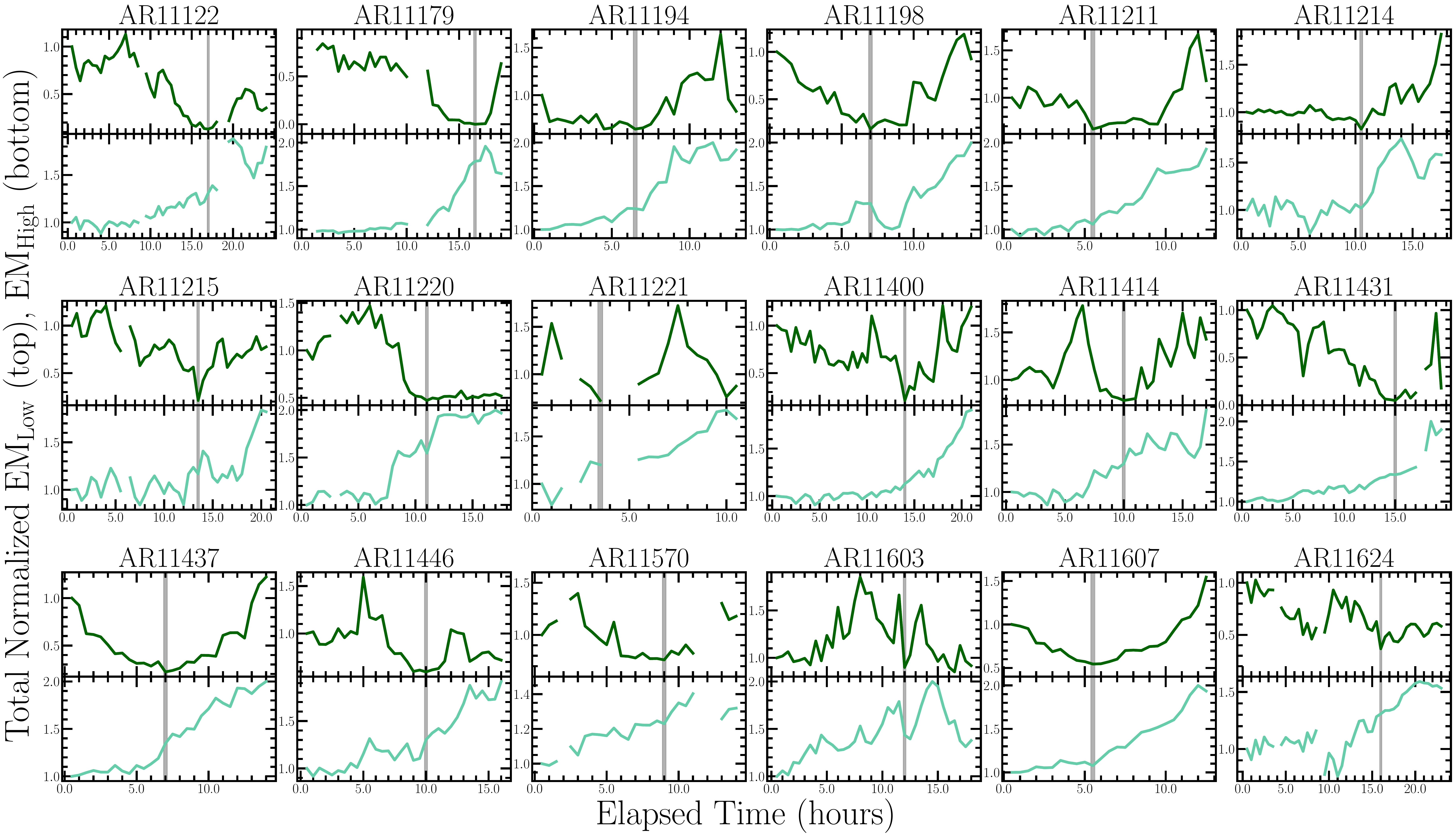}
    \caption{Spatially integrated EM for all ARs shown in Table \ref{table:dimmingtable}. For each AR, the top panel shows the normalized $\mathrm{EM_{low}}$ evolution and the bottom panel shows the normalized $\mathrm{EM_{high}}$ evolution. The transparent vertical line denotes the time of maximum dimming. The gap in the curves is due to missing AIA data. Error bars are too small to be visible.}
    \label{fig:all_dem_evolution}
    \vspace{4mm}
\end{figure*}


\begin{figure}[t!]
\centerline{\includegraphics[width=0.9\columnwidth]{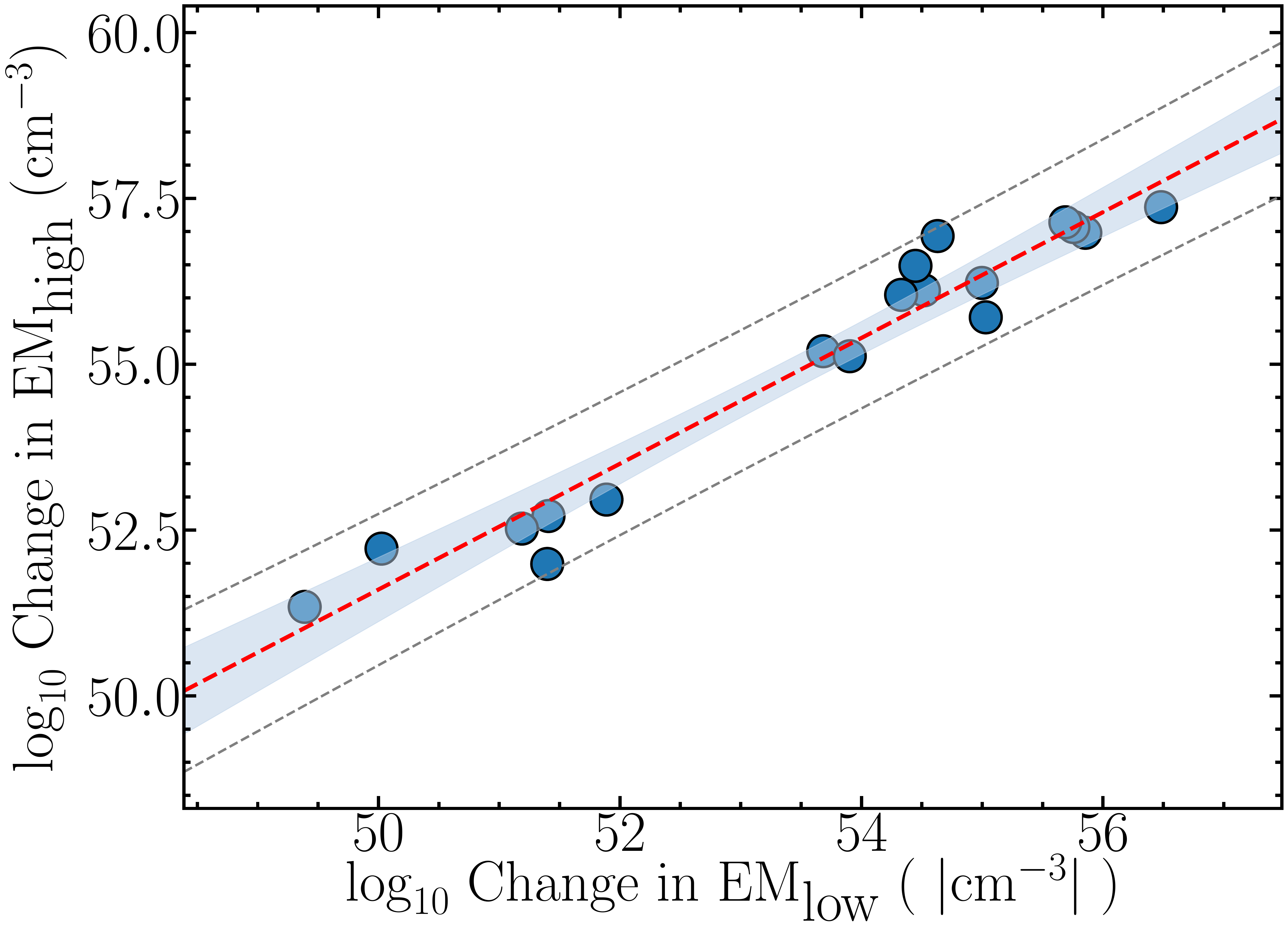}}
\caption{Scatter plot of the absolute change of the spatially integrated EM, $\left|\sum\mathrm{EM}(t_{*}) - \sum\mathrm{EM}(t_0)\right|$, in the low and high temperature bins for all events. Line shows the result of linear regression in the logarithm space. Error bars are too small to be visible. The best fit linear regression line, shown as the red dotted line, returns a slope of $0.95\pm0.12$ in log-log space corresponding to $r^2 =0.95$. The $95\%$ confidence limits and prediction limits are shown by the transparent region and dashed gray lines, respectively.   }
\label{fig:dem_scatter}
\end{figure}


\section{Results}
\label{sec:results}

The \textit{SDO} observations shown in Figure~\ref{fig:sdoaia} reveal several interesting features. The emerging dimming region resides to the west of the emerging flux with a fan-shaped boundary. In the AIA 171~{\AA} channel, the mean intensity in the emerging dimming region at maximum dimming $I(t_{*})$ is not only darker than the pre-dimming counterpart, but also much darker than the surrounding quiet Sun (QS) $I_\mathrm{QS}$, with $I(t_{*})/I_\mathrm{QS}=0.73$. In the AIA 211~{\AA} channel, the same region becomes brighter with $I(t_{*})/I_\mathrm{QS}=1.94$. Diffuse loop-like structure appear to connect the emerging dimming region to the emerging flux.

An example of DEM solution is shown in Figure \ref{fig:dem}. For this pixel, the DEM has a single peak at $\log_{10}T=6.2$ prior to the start of the dimming. At maximum dimming, however, the DEM peak shifts to $\log_{10}T=6.3$, and the peak value significantly increases. The DEM values in the range $5.7 \le \log_{10}T \le 5.9$, i.e., sub-MK coronal plasma typical for QS, decreases drastically during this period to near depletion. At the same time, the DEM values in the range $6.1 \le \log_{10}T \le 6.3$ increases by over 100$\%$. The amount of the plasma over 1~MK has significantly increased. Such behavior is expected from the AIA observations.

Figure~\ref{fig:dem_compare} shows the evolution of the spatially integrated EM in AR 11570. $\sum\mathrm{EM_{low}}$ increases first, then decreases until the maximum dimming is reached. $\sum\mathrm{EM_{high}}$, on the other hand, continues to increase throughout. The ratio between the total EM at the maximum dimming and pre-dimming times $\sum\mathrm{EM}(t_{*})/\sum\mathrm{EM}(t_0)$ are 0.27 and 1.38 for low- and high-temperature bins, respectively. The maximum dimming occurs 9~hr after the initial flux emergence.


\begin{figure}[t!]
\centerline{\includegraphics[width=0.8\columnwidth]{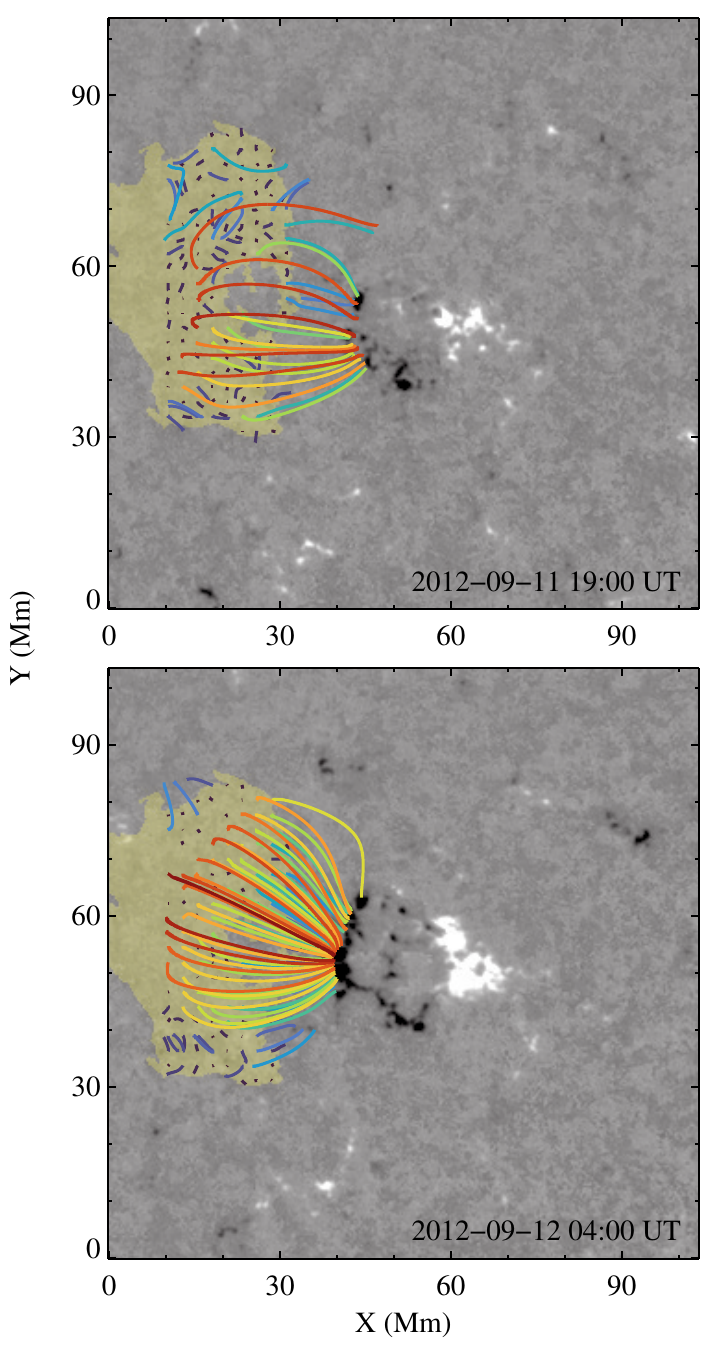}}
\caption{Top view of selective field lines from the potential field model for AR 11570. The two panels correspond to the two columns in Figure~\ref{fig:sdoaia}. Field lines in two panels have the identical starting foot points inside the dimming contours (shaded yellow). Colors indicate their lengths, with blue (red) being shorter (longer). The background image shows $B_r$ saturated at $\pm$500~G. }
\label{fig:ffmodel_combine}
\end{figure}


In our sample, the EM variation in the low-temperature channel exhibits significant variations in time, whereas the high-temperature channel increases more smoothly (Figure~\ref{fig:all_dem_evolution}).  We summarize in Table~\ref{table:dimmingtable} selective variables to quantify these evolution. Several trends are obvious.

\begin{enumerate}[noitemsep,topsep=0pt,parsep=2pt,partopsep=2pt,leftmargin=16pt]

\item The duration of emerging dimming, defined as the time difference between the maximum and the start of dimming $t_{*}-t_0$, is, on average, 10.9 hours.

\item At the maximum dimming, the AIA 171~{\AA} emission in the emerging dimming region is lower than the nearby QS, whereas in 211~{\AA} it is higher.

\item The change of total EM, $\sum\mathrm{EM}(t_{*}) - \sum\mathrm{EM}(t_0)$, is always negative (positive) for the low-$T$ (high-$T$) bin, as expected from our case selection criteria. The relative change, $1-\sum\mathrm{EM}(t_{*})/\sum\mathrm{EM}(t_0)$, can be over $95\%$ in the low-$T$ channel in the most extreme cases (AR 11179, AR 11431).

\item The total EM and its change are typically 1 to 2 orders of magnitudes greater in the high-$T$ channel. It may be partially due to the larger integration range in the real $T$ space. A bin size of 0.2 in $\log_{10}T$ translates to $\Delta T = 0.3$ and $0.9$~MK for the low- and high-$T$ channels, respectively.

\item Similar to \citet{zhang2012}, most events here display a fan-shaped morphology. Only 1 of the 9 events has a halo morphology.

\end{enumerate}

We find a strong correlation between the changes of total EM in the two temperature bins, as illustrated in Figure~\ref{fig:dem_scatter}. The EM values cover a wide range of 8 orders of magnitude. A best-fit linear regression in logarithm space reveals a slope of $0.95\pm0.12$.


\begin{figure}[t!]
\centerline{\includegraphics[width=0.86\columnwidth]{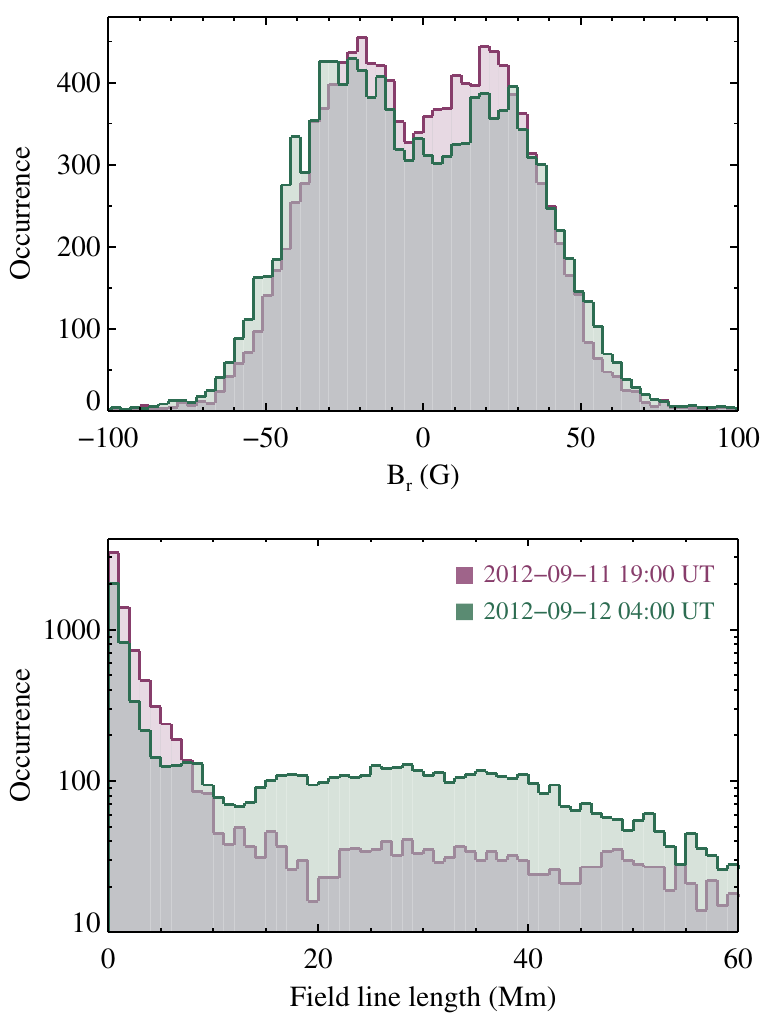}}
\caption{Top: histograms of the vertical magnetic field $B_r$ inside the dimming region in AR11570, at two times shown in Figure~\ref{fig:ffmodel_combine}. Bottom: histograms of the length of the field lines originating from the dimming region.}
\label{fig:hist_t1}
\end{figure}


Figure~\ref{fig:ffmodel_combine} shows selective potential field lines originated within the dimming contour for AR 11570. Figure~\ref{fig:hist_t1} shows the histograms of $B_r$ and the field line lengths before and during emerging dimming. While the magnetic field within the contour does not appear to change much (Figure~\ref{fig:sdoaia} and Figure~\ref{fig:hist_t1}), the local magnetic connectivity does change significantly over the emerging dimming period. Many field lines that originally close locally now appear to connect to the newly emerged flux via longer, higher-arching loops. This is consistent with the loop-like features observed in AIA 211~{\AA}.


\section{Discussion \& Conclusion}
\label{sec:discussion}

We have analyzed a total of 18 emerging dimming events in a quantitative manner to understand their physical origin. Our DEM analysis shows a marked change of the total EM in the emerging dimming regions. The simultaneous decrease in the range $5.7 \le \log_{10}T \le 5.9$ and increase in $6.2 \le \log_{10}T \le 6.4$ well explain the behaviors observed in AIA 171 and 211~{\AA} channels.

We find two lines of evidence in support of the hypothesis that emerging dimming is caused by coronal heating. First, the changes of $\sum\mathrm{EM}$ in the two temperature bins are well correlated. The fact that such correlation holds over 8 orders of magnitude strongly suggest a common physical cause. As the heating acts on plasma of all temperatures, a proportionality between different bins are perhaps not surprising. Another common mechanism for EUV dimming, mass loss via plasma ejection or expansion \citep[e.g.,][]{harra2001}, will result in a decrease of EM in all temperature bins and can thus be ruled out. Second, the change of magnetic connectivity suggests ongoing reconnection, which provides a viable energy source for coronal heating. Such heating episodes due to new flux emergence have been quantitatively studied before \citep{tarr2014}. Additionally, the emerging dimming region magnetic field remains QS-like. We find that the net magnetic flux is well balanced. Most loops are closed, with very few open field lines despite the small computation domain. A mass-loss scenario is thus unlikely.

Our sample size is relatively small, and the selection criteria are somewhat arbitrary (for example, no event is included between June 2011 and January 2012). Nevertheless, the fact that all events studied evolve in a similar fashion suggests that our conclusion is not biased by the sample selection. We further note that the two parent samples our study is based on, i.e., the emerging dimming sample \citep{zhang2012} and the flux emergence sample \citep{schunker2016}, are both complete.

We discuss two interesting aspects mentioned in \citet{zhang2012} and reproduced in this study. First, there is a delay of hours between the start of flux emergence and the maximum dimming. Both the flux emergence rate and the AR total flux were shown to negatively correlate with the delay, and positively correlate with the emerging dimming duration. Such is consistent with a magnetic reconnection origin. Second, most emerging dimming events have a fan-like morphology rather than a halo. Such is likely determined by the property of the ambient field. If it is mostly unipolar (as in a coronal hole), the minority polarity of the emerging bipole will reconnect in all directions. This naturally leads to a dome-shaped separatrix \citep[e.g.,][]{tarr2014}, whose quasi-circular footprints map to the halo dimming region.

A recent study found reduced 171~{\AA} emission in moat-like regions around seven isolated, well-formed ARs \citep{singh2021}. These extended dark moats are also visible in the 304, 131, 94, and 335~{\AA} channels, and less so in the 193 and 211~{\AA} channels. There are no signs of brightening. A DEM analysis indicates a reduction of plasma in the entire $5.7 \le \log_{10}T \le 6.2$ range. These observations point to a physical origin different from the emerging dimming. Following \citet{antiochos1986}, the authors propose that the magnetic loops from the moats are pushed to low altitudes by the strong AR fields, and are thus restricted to lower-than-coronal temperatures.

Similar dimming is reported for Sun-as-a-star synthetic observations when a large, isolated AR is transiting the disc \citep{toriumi2020}. The EUV irradiance in all AIA channels are positively correlated with the sunspot magnetic flux except for 171~{\AA}. The AR studied is relatively stable with no significant flux emergence. Hence, the cause may be more in line with that of \citet{singh2021}. If it is related to emerging dimming, it is possible that the depleted low-temperature plasma is never replenished. For more active stars with larger stellar spots, the effect may be more significant.

We note that magnetic flux emergence simulations can now reproduce realistically many observed structures on AR scales \cite[e.g.,][]{toriumi2019}. When extended to include a coronal domain \cite[e.g.,][]{cheung2019} or coupled with a coronal model, they will provide a self-consistent ``digital laboratory'' for testing the relevant physics for emerging dimming.

We conclude that emerging dimming is likely related to coronal heating episodes powered by reconnection between the emerging and the ambient magnetic fields. An assessment of the mass and energy budget will be deferred to future investigation.


\begin{acknowledgments}
This work is supported by the state of Hawai`i and NSF award \#1848250. We thank M. Cheung and W. Liu for discussion and help on the DEM error analysis. The \textit{SDO} data are courtesy NASA, the HMI, and the AIA science teams.
\end{acknowledgments}

\facilities{\textit{SDO}}


\end{CJK*}



\end{document}